\documentclass[acmtog]{acmart}

\usepackage{booktabs} %

\setcopyright{acmcopyright}
\acmJournal{TOG}
\acmYear{2020}
\acmVolume{39}
\acmNumber{6}
\acmArticle{195}
\acmMonth{12}
\acmDOI{10.1145/3414685.3417815}

\usepackage{amsmath,amsthm,amssymb}
\usepackage{graphicx}
\usepackage{textcomp}
\usepackage{wrapfig}
\usepackage{subfig}
\usepackage{color}
\usepackage{xspace}
\usepackage{overpic}
\usepackage{subfig}
\usepackage{enumitem}
\usepackage{booktabs}

\usepackage{color}
\definecolor{blue}{rgb}{0,0,1}
\definecolor{red}{rgb}{1,0,0}
\definecolor{green}{rgb}{0,.5,0}
\definecolor{orange}{rgb}{0.75, 0.4, 0}
\newcommand{\bj}[1]{{\color{black}\textbf{}#1}\normalfont}

\newcommand{\etal}{et al.}

\begin{document}

\title[Differentiable Refraction-Tracing for Mesh Reconstruction of Transparent Objects]{Differentiable Refraction-Tracing for Mesh Reconstruction of Transparent Objects}

\author{Jiahui Lyu}
\authornote{Equal contribution}
\affiliation{%
	\institution{Shenzhen University}
}
\author{Bojian Wu}
\authornotemark[1]
\affiliation{%
	\institution{Alibaba Group}
}
\author{Dani Lischinski}
\affiliation{%
	\institution{The Hebrew University of Jerusalem}
}
\author{Daniel Cohen-Or}
\affiliation{
	\institution{Shenzhen University}
}
\affiliation{
	\institution{Tel Aviv University}
}
\author{Hui Huang}
\authornote{Corresponding author: Hui Huang (hhzhiyan@gmail.com)}
\affiliation{%
	\department{College of Computer Science \& Software Engineering}
	\institution{Shenzhen University}
}

\renewcommand\shortauthors{J. Lyu, B. Wu, D. Lischinski, D. Cohen-Or, and H. Huang}

\begin{abstract}
Capturing the 3D geometry of transparent objects is a challenging task, ill-suited for general-purpose scanning and reconstruction techniques, since these cannot handle specular light transport phenomena. Existing state-of-the-art methods, designed specifically for this task, either involve a complex setup to reconstruct complete refractive ray paths, or leverage a data-driven approach based on synthetic training data. In either case, the reconstructed 3D models suffer from over-smoothing and loss of fine detail. This paper introduces a novel, high precision, 3D acquisition and reconstruction method for solid transparent objects. Using a static background with a coded pattern, we establish a mapping between the camera view rays and locations on the background. Differentiable tracing of refractive ray paths is then used to directly optimize a 3D mesh approximation of the object, while simultaneously ensuring silhouette consistency and smoothness. Extensive experiments and comparisons demonstrate the superior accuracy of our method.
\end{abstract}



\begin{teaserfigure}
	\centering
	\includegraphics[width=1.0\linewidth]{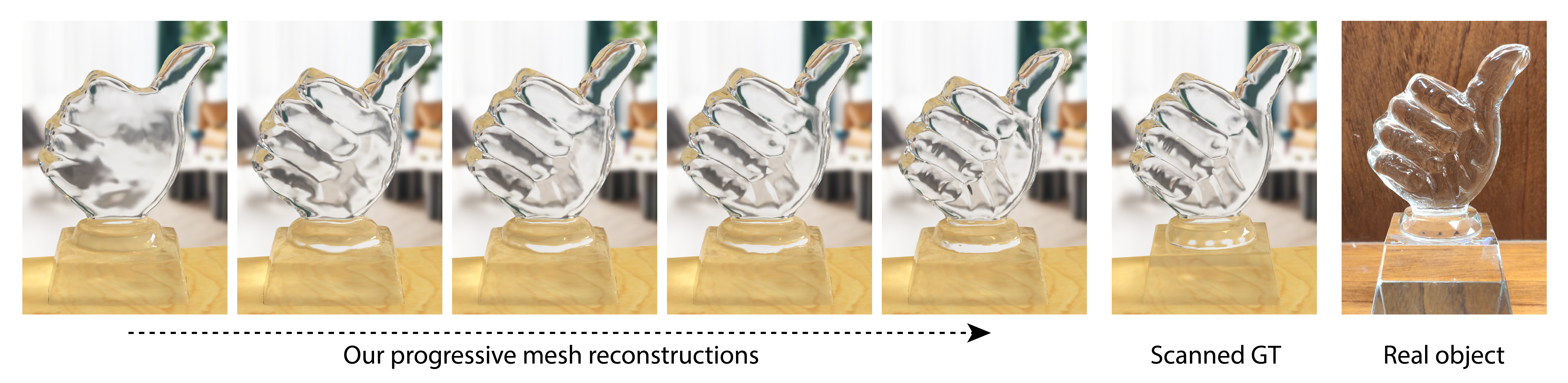}
	\caption{Reconstructing a transparent \textit{Hand} object. The five images, from left to right, show a sequence of ray-traced models, progressively optimized by our method. The ground-truth geometry, obtained by painting and scanning the object and a real photograph of the original object are shown on the right.}
	\label{fig:teaser}
\end{teaserfigure}

\maketitle

\section{Introduction}
\label{sec:intro}

Acquiring the 3D geometry of real world objects has been one of the longstanding problems in the fields of computer graphics and computer vision. Most existing 3D acquisition approaches, such as laser scanning and multi-view reconstruction, are based on the assumption that the object is opaque and its surface is approximately Lambertian. Thus, such approaches are not applicable to objects made from a transparent, refractive material, due to the complex manner in which they interact with light.

Several methods have recently been proposed for non-intrusive acquisition of such objects~\cite{wu2018frt,li2020through}. The method of Wu et al.~\shortcite{wu2018frt} is based on capturing correspondences between camera rays and the rays incident on the object from behind. This necessitates a complex setup with a rotating background monitor, and yields a point cloud from which the final model is consolidated. Li et al.~\shortcite{li2020through} employ a data-driven approach, which leverages a large number of synthetic images as its training set, and requires capturing the environment map. Both of these methods are limited in their ability to recover fine geometric detail, resulting in overly smooth reconstruction results, as we demonstrate in Section~\ref{sec:results}.

In this work, we propose a novel, non-intrusive method for reconstructing a detailed 3D mesh of solid transparent objects. In contrast to Wu et al.~\shortcite{wu2018frt}, our approach is based on optimizing correspondences between camera rays and locations on a static background monitor, thereby cutting the acquisition time by half, \bj{and avoiding additional cumulative errors}. More importantly, the proposed method optimizes the reconstructed mesh directly, and is able to capture the fine geometric details of the object's surface, as may be seen in Fig.~\ref{fig:teaser}. \bj{Furthermore, our approach leverages automatic differentiation, which can be better integrated with popular deep learning frameworks and benefit from GPU-accelerated optimization.}

Starting from a rough initial mesh, obtained from the visual hull of the object, our method progressively refines the mesh in a coarse-to-fine fashion, as shown in Figure~\ref{fig:teaser}. Specifically, via differentiable tracing of refracted ray paths, our method optimizes an objective function that consists of three losses:
\begin{enumerate}
	\item Refraction loss, which minimizes the distance between the observed background refractions and the simulated ones;
	\item Silhouette loss, which ensures that the boundary of the optimized mesh matches the captured silhouettes;
	\item Smoothness loss, ensuring smoothness of the optimized mesh.
\end{enumerate}

\bj{
Our approach only makes use of refractive ray paths through the object that feature exactly two refractions, once upon entering, and once upon exiting the transparent object. Thus, our optimization ignores some of the additional ray paths that may be observed during acquisition, i.e., those involving more than two intersections with the object's surface and/or total internal reflection. The effect of these assumptions is discussed in Section \ref{subsec:discussion}. %
}

In the remainder of this paper, we first briefly survey the related previous work.
Following an overview of our approach (Section~\ref{sec:overview}), we describe the terms of our objective function in more detail (Section~\ref{sec:recon}). Results and comparisons with previous work \cite{wu2018frt,li2020through} are presented in Section~\ref{sec:results}. Section~\ref{sec:conclusion} concludes the paper and suggests directions for future work.

\section{Related Work}\label{sec:related}

\subsection{Environment Matting}

Matting is a process concerned with extracting from an image a scalar foreground opacity map, commonly referred to as the alpha channel~\cite{Porter:1984,Levin:2008}. Environment matting is an extension of alpha matting that also captures how a transparent foreground object distorts its background, so it may be composited over a new one.
The pioneering work of Zongker \etal~\shortcite{zongker1999environment} extracts the environment matte from a series of projected horizontal and vertical stripe patterns, with the assumption that each pixel is only related to a rectangular background region.
To improve environment matting accuracy and to better approximate real-world scenarios, Chuang~\etal~\shortcite{Chuang:2000:EME} propose to locate multiple contributing sources from surrounding environments. Other works present solutions in domains other than the image, such as the wavelet domain~\cite{peers2003wavelet}, or the frequency domain~\cite{qian:2015:envmat}. 
Our approach could be viewed as an extension of environment matting to the task of transparent object reconstruction, in the sense that it progressively optimizes the reconstructed shape of the object so as to better match a collection of environment mattes captured from multiple views.

\subsection{Transparent surface reconstruction}

Reconstructing the surface geometry of transparent objects is a longstanding challenging problem~\cite{ihrke2010transparent}. Some methods use destructive or intrusive techniques~\cite{hullin2008fluorescent,Aberman:2017:DTS:3072959.3073693,trifonov2006tomographic} to obtain a detailed surface geometry. Non-intrusive methods use the refractive properties of transparent objects to recover their shape by analyzing the distortions of reference background images~\cite{Ben-Ezra:2003:MRT:946247.946706,wetzstein2011refractive,tanaka2016recovering}.
Recovering the object shape from the optical distortion it induces is typically applicable to a single refractive surface or a parametric model, since light transport resulting from multiple reflections and refractions is much more difficult to analyze.
In addition to intrinsic refractions, it is also possible
to capture the reflective components of light transport, and estimate the shape geometry by observing exterior specular highlights~\cite{morris2007reconstructing,yeung2011adequate}. Since reflections occur on the outermost surface, it is possible to reconstruct objects with complex geometries and inhomogeneous internal materials. However, the acquisition process is quite involved and considerable manual effort is needed to control the lighting conditions precisely enough to obtain reasonable results.

More recently, several researchers tackle the task of transparent object reconstruction by incorporating deep learning techniques. Stets \etal~\shortcite{stets2019single} and Sajjan \etal~\shortcite{sajjan2020cleargrasp} propose to use encoder-decoder architectures for estimating the segmentation mask, depth map and surface normals from a single input image of a transparent object. Li~\etal~\shortcite{li2020through} present a different approach, where a rendering layer is embedded in the network to account for complex light transport behaviors. They achieve state-of-the-art reconstruction results using multi-view images. Due to the difficulty of obtaining a sufficient amount of real training data, these data-driven methods rely on synthetic training images. Although these images are generated using high-fidelity photorealistic rendering, the domain gap between real-world and synthetic images still exists.
Specifically, networks trained on synthetic images have difficulty generalizing to real input images, and thus they are prone to reconstruction errors, as will be demonstrated in Section~\ref{sec:results}. \bj{In contrast, we use a controlled acquisition setup to capture refractive light paths and use direct per-object shape optimization, which does not require a training set consisting of similar shapes.}

\subsection{Light path triangulation}

Light path triangulation is an extension of classical stereo triangulation, which uses the relationship between direction of refraction and the surface normal to infer geometry from light transport.
Kutulakos and Steger~\shortcite{kutulakos2008theory} provide a theoretical analysis of the reconstruction feasibility based on the number of specular reflections and refractions along the ray paths.
The simplest case involves a single refraction~\cite{shan2012refractive}, such as when reconstructing a water surface~\cite{morris2011dynamic,Qian_2017_CVPR,zhang2014recovering}. Next, Tsai~\etal~\shortcite{tsai2015does} reveal depth-normal ambiguity while assuming that the light rays refract twice. To eliminate the ambiguity, Qian~\etal~\shortcite{Qian_2017_CVPR} propose a position-normal consistency based optimization framework to recover front and back surface depth maps. Wu~\etal~\shortcite{wu2018frt} extend this approach and present the first non-intrusive method to reconstruct the full shape of a general transparent object; however, due to their separate optimization and multi-view fusion of recovered point clouds, the results are always over-smoothed. Following Wu \etal~\shortcite{wu2018frt}, we also start our optimization from the object's visual hull; however, rather than fusing point clouds, we directly optimize the surface mesh by leveraging differentiable rendering techniques. This approach enables us to recover fine-grained geometric detail, as demonstrated by the comparisons in Section~\ref{sec:results}. 

\bj{
\subsection{Differentiable rendering}
Multiple works utilize differentiable rendering for image-based 3D mesh reconstruction, such as Neural 3D Mesh Renderer~\cite{kato2018neural} and Soft Rasterizer~\cite{liu2019soft}. These methods typically assume a simplified image formation model and are limited to Lambertian scenes. 
Li et al.~\shortcite{li2018dmc} introduce a general-purpose differentiable ray tracer that is able to compute derivatives of scalar functions over a rendered image with respect to arbitrary parameters.
For transparent objects, Nimier-David~\etal~\shortcite{nimier2019mitsuba} propose Mitsuba~2, a retargetable differentiable renderer, that could be applied in computational caustics design~\cite{papas2011goal}. Differently from these methods, we do not employ an irradiance-based loss function that measures the discrepancy between the pixel values of the rendered image and the ground truth. Rather, our refraction loss is based directly on ray-pixel correspondences, which reflect the geometry of the underlying light transport. The geometry of light paths is directly determined by the shape geometry, which is what we seek to recover, compared to the final RGB pixel colors, which are influenced by additional factors, such as the BRDF.
}
\section{Overview}\label{sec:overview}

\begin{figure}
	\centering
	\includegraphics[width=1.0\linewidth]{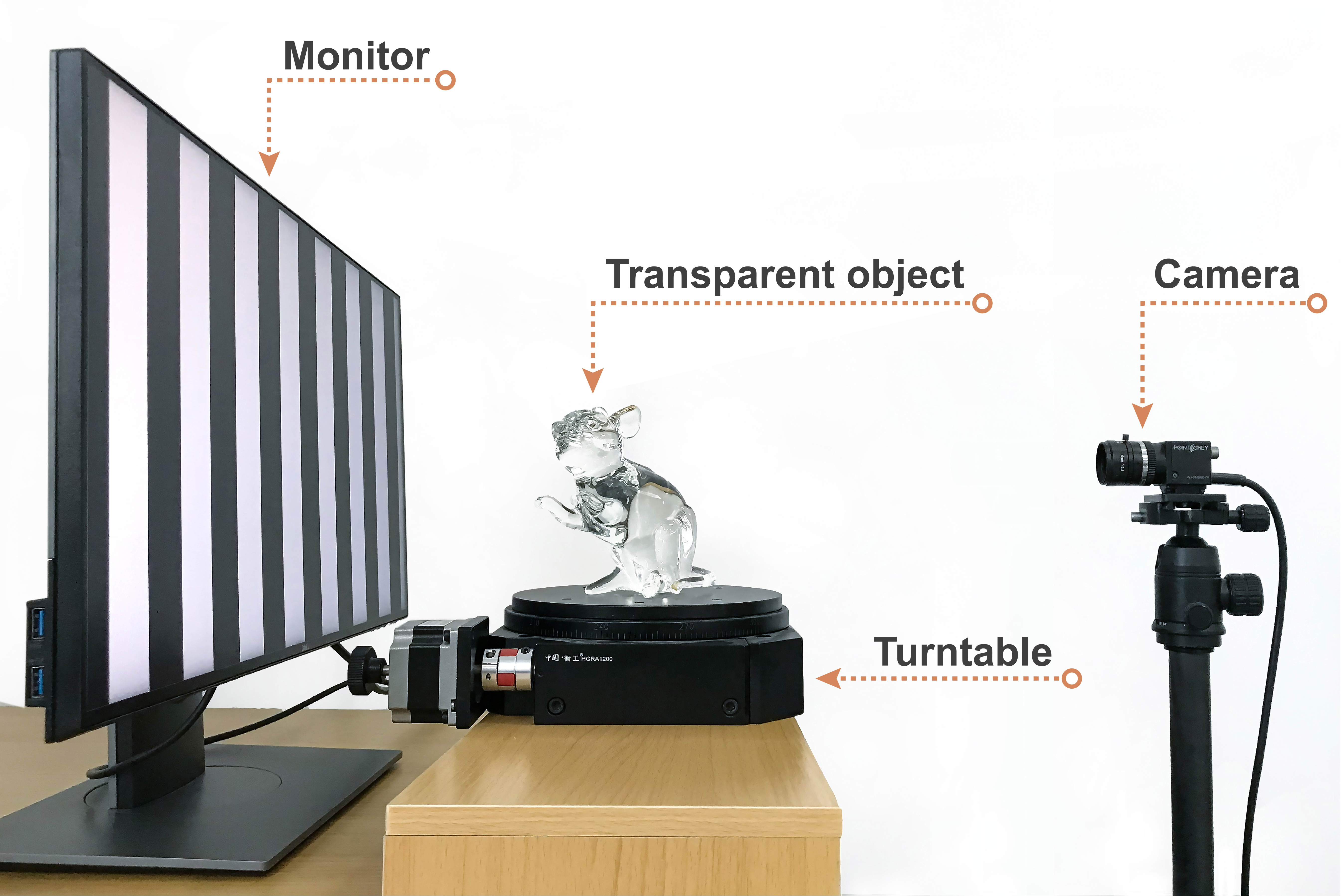}
	\caption{Our transparent object capture setup. The object to be captured is placed on the turntable, which is rotated during acquisition to provide the static camera with multiple views of the object. A static LCD monitor is placed behind the object, displaying horizontal and vertical stripe patterns that form a Gray-coded background. The background is used for extracting the object's silhouette and estimating the environment matte for each camera view.}
	\label{fig:setup}
\end{figure}

\begin{figure*}[t!]
	\centering
	\includegraphics[width=1.0\linewidth]{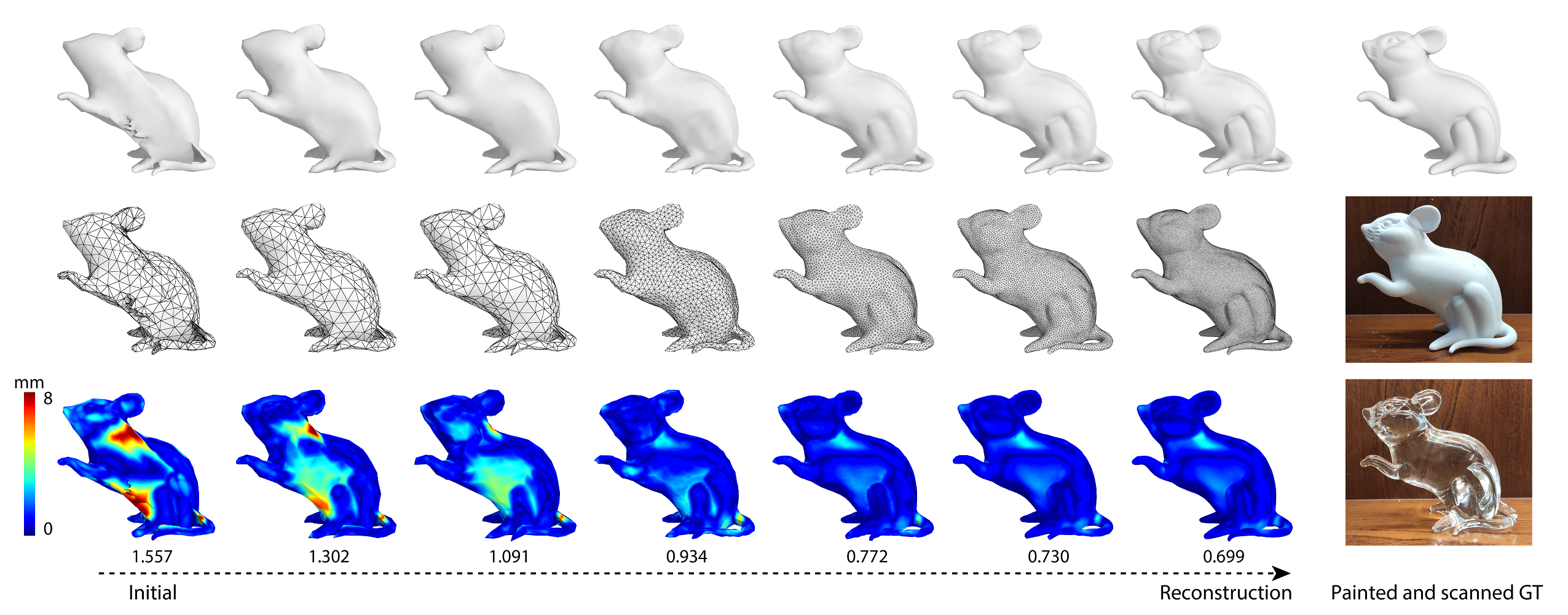}
	\caption{Coarse-to-fine reconstruction of a real \textit{mouse} statue. Top: starting with the visual hull obtained by space carving, our method gradually recovers details ranging from large geometric displacements such as neck and tummy to fine-level details like eyes. Middle: we alternatively remesh and reconstruct geometric detail at progressively finer scales. Bottom: the error map is visualized using the shortest distance between each vertex of the reconstruction and the ground truth mesh. The number below is the average of the per-vertex distances in millimeters. The real size of the statue's bounding box is $178\textit{mm} \times 101\textit{mm} \times 71\textit{mm}$.}
	\label{fig:lod}
\end{figure*}

\begin{figure}
	\centering
	\includegraphics[width=1.0\linewidth]{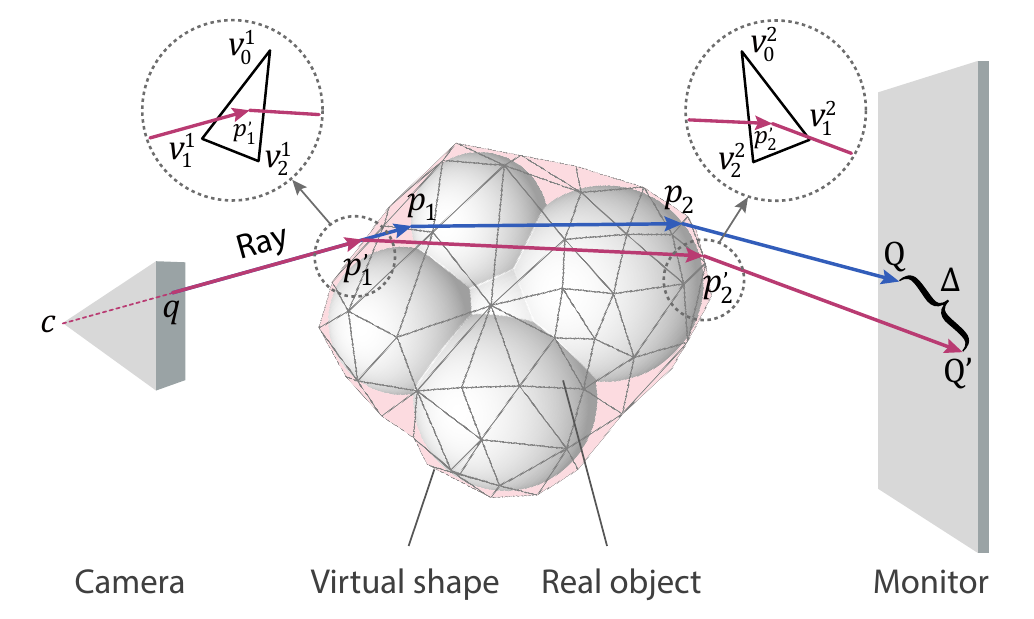}
	\caption{Refraction loss. The simulated refractive ray path (in red) through image pixel $q$ should reach the observed background point $Q$, which corresponds to the intersection of the real ray path (in blue) with the background monitor. The pink mesh is the optimized virtual shape, initialized to the visual hull. The top left and right insets show the associated triangles and vertices of a single simulated ray-pixel correspondence.}
	\label{fig:ray}
\end{figure}

Our approach for transparent object reconstruction may be viewed as \emph{reconstruction-by-synthesis}. 
We acquire multiple views of the target transparent object, with a Gray-coded background pattern refracted through the object in each view, and our goal is to reconstruct a virtual 3D model of the object, that would refract the background in a manner consistent with the captured observations. Our capturing setup is shown in Fig.~\ref{fig:setup}, and is further described in Section~\ref{subsec:setup}.

The reconstruction is carried out by a coarse-to-fine optimization process, visualized in Fig.~\ref{fig:lod}.
The optimization starts from a rough initial mesh, which is progressively remeshed and deformed to better match the observed background distortion and the silhouette of the object in the captured views.
The coarse stages of the optimization introduce large deformations of the optimized 3D mesh, while later stages introduce and refine finer geometric surface detail. 

The rough initial shape is obtained from the visual hull defined by the multiview silhouettes. The subsequent optimization is mainly driven by differentiable tracing of refractive ray paths. Specifically, our goal is to deform the virtual shape such that the ray paths refracted by it twice (upon entering and upon exiting) reach the same points on the background pattern that are visible in the captured views, as shown in Fig.~\ref{fig:ray}.
In other words, we know the background point $Q$ that corresponds to each pixel $q$ in each captured image, and use differentiable ray tracing to optimize the shape such that the corresponding ray path indeed reaches $Q$.
Formally, this goal is achieved by minimizing the \emph{refraction loss}; see Section~\ref{subsec:loss_refract}.

In addition to the refraction loss, our optimization also makes use of \emph{silhouette loss}, which ensures that the boundary of the optimized mesh matches the silhouettes, as seen from the different views.
Finally, a \emph{smoothness loss} term is used to ensure the smoothness of the resulting reconstructed mesh.

\section{Method}\label{sec:recon}

\newcommand{\Lref}{\mathcal{L}_{\textit{refract}}}
\newcommand{\Lsil}{\mathcal{L}_{\textit{silhouette}}}
\newcommand{\Lsm}{\mathcal{L}_{\textit{smooth}}}

\begin{figure*}[t!]
	\centering
	\includegraphics[width=1.0\linewidth]{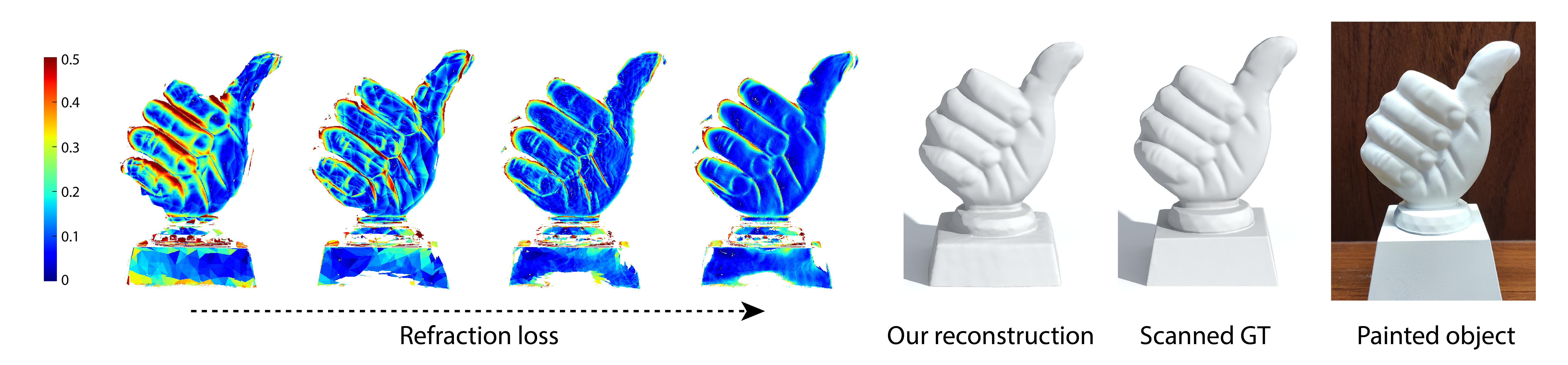}
	\caption{Visualization of the refraction loss, as defined in Eq.~\eqref{eq:loss_refract}, of valid rays involved in two refractions. The visualized loss values are normalized by the resolution of the background monitor. As the refraction loss decreases during optimization, the final reconstructed mesh closely approximates the scanned ground truth, and correctly reproduces light transport, as shown in Fig.~\ref{fig:teaser}.}
\label{fig:hand}
\end{figure*}

As previously mentioned, our method starts by reconstructing an initial rough shape of the object from a collection of multiview silhouettes using space carving~\cite{SpaceCarving}. Subsequently, the rough model is progressively optimized, based on the correspondences between view rays and background locations, extracted by environment matting~\cite{zongker1999environment}, while maintaining the boundary constraints, provided by the silhouettes. In order to recover geometric details at progressively finer scales, we alternate between remeshing the shape and optimizing the loss function in a coarse-to-fine fashion to reach the goal; see Fig.~\ref{fig:lod} for example. At each stage, the reconstructed shape is gradually updated by minimizing a combination of three terms: refraction loss, silhouette loss and smoothness loss:
\begin{equation}\label{eq:loss_function}
\centering
\mathcal{L} = \alpha\Lref + \beta\Lsil + \gamma\Lsm,
\end{equation}
where $\alpha$, $\beta$ and $\gamma$ are balancing coefficients for the loss terms, which we set to $10^4/HW$, $0.5/\min(H,W)$ and $10^3 / \texttt{edgelen}$ by default, respectively, where $H$ and $W$ represent the height and width of the camera imaging plane, and $\texttt{edgelen}$ denotes the average edge length in the currently optimized mesh. In the following subsections, we describe each of our loss terms in more detail.

\subsection{Refraction loss}\label{subsec:loss_refract}

As shown in Fig.~\ref{fig:ray}, given the current virtual shape mesh, we first trace rays from the camera as they intersect and refract through the shape, and then optimize the positions of associated mesh vertices according to the captured correspondences between view rays and background locations (e.g., the ray $cq$ and point $Q$ in Fig.~\ref{fig:ray}). The optimization only makes use of rays that refract exactly twice before reaching the background, first when the camera ray enters the virtual shape ($p'_1$), and once more when it exits the shape ($p'_2$). The full simulated light path, shown in red in Fig.~\ref{fig:ray}, intersects the background at $Q^\prime$. Since $Q^\prime$ is generally different from the destination of the actual optical path through the real object, which is shown in blue and terminates at $Q$, our goal is to minimize $\Delta = \left|\left| Q - Q^\prime \right|\right|^2$. 

\bj{Note that while our approach only utilizes ray paths with two refractions, this limitation is actually less restrictive than it may seem. In particular, it does not limit our approach to convex objects, as demonstrated by all of our examples in this paper. The reason is that each reconstructed object is captured from multiple view directions, and as the number of views increases, most mesh triangles end up having two-refraction ray paths passing through them. The effect of considering only two-refraction paths is further discussed and demonstrated in Section~\ref{subsec:discussion}.}

In comparison, Wu \etal~\shortcite{wu2018frt} also optimize the recovered shape based on captured environment mattes; however, their approach requires moving the background for each view to reconstruct the 3D rays exiting the object towards the background, and yields point clouds. In contrast, we rely on a single environment matte per view, extracted using a static background, and leverage differentiable ray tracing to directly optimize the reconstructed 3D mesh.

\begin{figure*}[t!]
	\centering
	\includegraphics[width=1.0\linewidth]{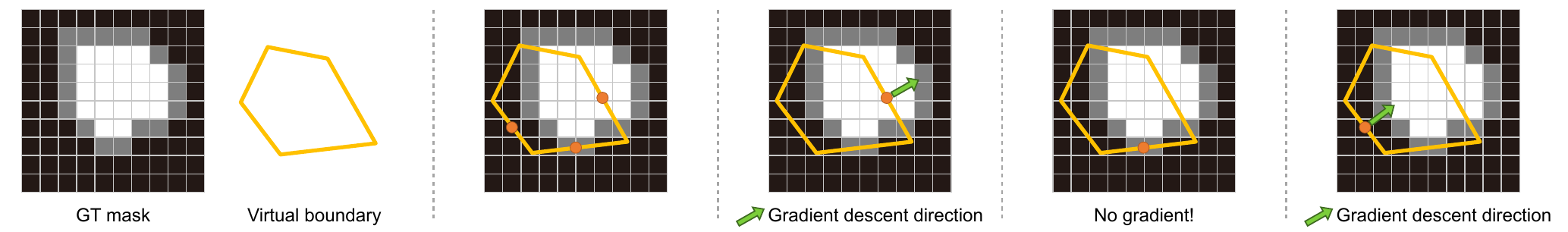}
	\caption{Silhouette loss and its negative gradient direction. For each captured camera view, the ground truth mask of the object encodes pixels inside, outside, and on the boundary of the object (white, black, and grey, respectively). The projected silhouette edges of the reconstructed mesh are shown in yellow. The gradient is set to zero for edges whose center point falls on the mask boundary, and is non-zero for edges inside or outside the mask.}
	\label{fig:loss_sil}
\end{figure*}

As mentioned above, our goal is to minimize the gap between simulated background position ($Q^\prime$) and its corresponding captured ground truth ($Q$), in order to cause the simulated and real light paths to coincide. Specifically, since each simulated ray path is associated with two triangles that it intersects, the vertices of these triangles are optimized to achieve this goal. The whole process is accomplished by iterative ray tracing and optimization, using the loss function defined as follows:
\begin{equation}\label{eq:loss_refract}
\centering
\Lref = \sum_{u=1}^U \left( \sum_{i \in I} \left|\left| Q - Q^\prime \right|\right| ^2 \right),
\end{equation}
where $U$ is the number of captured views and the set $I$ only contains ray paths that go through the object and involve exactly two refractions.
Let $v_k^1$ and $v_k^2$ ($k=0,1,2$) represent vertices of triangles that contain $p_1^\prime$ and $p_2^\prime$, respectively, as shown in Fig.~\ref{fig:ray}. Since the normal of each triangle is uniquely determined by its vertices, and the refractions are governed by Snell's law~\cite{born2013principles}, the final exiting ray $p_2^{\prime}Q^{\prime}$ is fully parameterized by $v_k^1$ and $v_k^2$. Thus, $Q^\prime$, obtained by intersecting the ray with the background monitor, is also a function of the associated vertices. Since all the operations to obtain $Q'$ are differentiable, the gradient of Eq.~\eqref{eq:loss_refract} is easily calculated using the chain rule. The effect of the refraction loss is visualized in Fig.~\ref{fig:hand}.

\begin{figure*}[t!]
	\centering
	\includegraphics[width=1.0\linewidth]{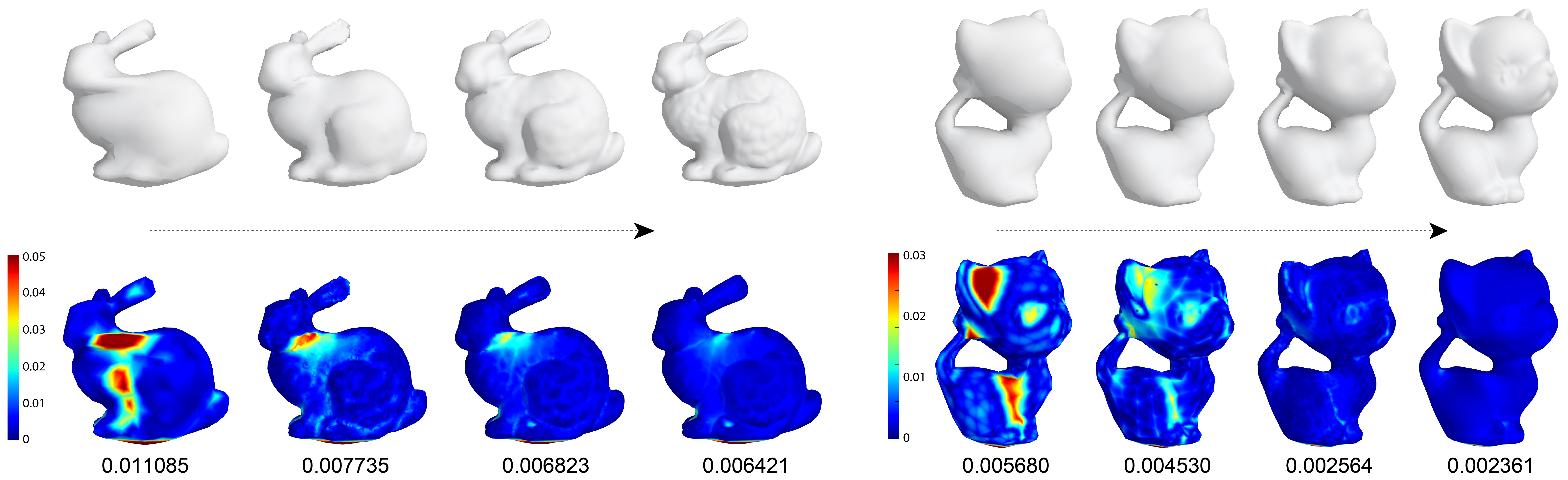}
	\caption{Coarse-to-fine reconstruction of synthetic \textit{Bunny} (left) and \textit{Kitten} (right) models, rendered with a refractive index set to 1.5. The top row shows the progression from an initial shape obtained by space carving to a more detailed reconstruction. The bottom row visualizes the per-vertex reconstruction error, measured by the minimal distance between each vertex in the top row models and the ground truth. The average distance, below each visualization, is decreasing, as expected. Each of the two models is scaled such that the longest dimension of its bounding box is 1.0.} 
	\label{fig:bunny_kitten}
\end{figure*}

\subsection{Silhouette loss}

The silhouettes of the object in the captured views offer an important set of constraints for the boundary of the reconstructed virtual shape. Although the initial visual hull, obtained by space carving, perfectly matches the captured silhouettes, optimizing using the refraction loss alone might cause the silhouettes of the virtual shape to deviate from the ground truth ones, as demonstrated in Section~\ref{subsec:ablation}. The reason for this is that there may exist multiple refractive surfaces that might satisfy the observed refraction of the background pattern. Thus, we introduce $\Lsil$, a novel loss term, to ensure that the silhouettes of the reconstructed shape are consistent with the captured silhouettes.

Specifically, the silhouette loss is defined using the ground truth projection mask of the object, as captured for each view, and the projection of the silhouette edges of the virtual shape, as seen from the same view (see Fig.~\ref{fig:loss_sil}). Let $b$ denote the projection of a virtual silhouette edge onto the image plane, and $s_b$ denote the midpoint of that edge, which serves as a representative of the edge for the computation of the silhouette loss. The loss is defined as the number of edge midpoints strictly outside or strictly inside the ground truth mask:
\begin{equation}
	\centering
	\Lsil = \sum_{u=1}^{U}\sum_{b \in B}\left|\chi(s_{b})\right|.
	\label{eq:loss_sil}
\end{equation}
Here, $\chi(s)$ is an indicator function whose value is $1$ or $-1$ if $s$ is inside or outside of the ground truth mask, and $\chi(s) = 0$ when $s$ lies on the mask boundary. $B$ is the collection of the virtual silhouette edges, obtained by projecting their 3D counterparts onto the image plane of each view. Since $\chi(s)$ is a discrete function, we define its gradient manually. Specifically, the shape of the virtual mesh is used to determine the direction of gradient descent (negative gradient). Let $N_b$ denote the normal of edge $b$, which points to the outside of the virtual boundary. As shown in Fig.~\ref{fig:loss_sil}, if $s_b$ is inside the ground truth mask, $b$ should move along $N_b$ in order to align with the ground truth. Conversely, if $s_b$ is outside the mask, it should move along $-N_b$. Finally, if $s_b$ is on the mask boundary, its gradient is set to zero. In summary, the negative gradient is defined as follows:
\begin{equation}
	\centering
	-\nabla \left|\chi(s_{b})\right| := \chi(s_b)\left|\left| b \right|\right| N_b,
	\label{eq:loss_sil_grad}
\end{equation}
where $\left|\left|b\right|\right|$ denotes the length of $b$. Let $b$ be the image projection of the 3D mesh edge between the vertices $v^b_1$ and $v^b_2$. Then the projected edge midpoint $s_b$ is given by:
\begin{equation}
	\centering
	s_b = \mathbf{P}_u \frac{v^b_1 + v^b_2}{2}
	\label{eq:proj},
\end{equation}
where $\mathbf{P}_u$ is the projection matrix of view $u$. Thus, the gradient of Eq.~\eqref{eq:loss_sil} with respect to the 3D mesh vertices can be calculated by combining Eq.~\eqref{eq:loss_sil_grad} and the gradient of Eq.~\eqref{eq:proj} via the chain rule.

\subsection{Smoothness loss}

Finally, we incorporate an additional loss term to encourage the reconstructed mesh to be smooth during the optimization. This term measures the discrepancy between normal vectors of neighboring triangles:
\begin{equation}\label{eq:loss_smooth}
	\centering
	\Lsm = \sum_{e \in E} \left( -\log(1 + \langle N_1^e, N_2^e\rangle) \right),
\end{equation}
where $E$ is the set of all the edges, and $\langle N_1^e, N_2^e\rangle$ is the dot product of the normals of the two adjacent triangles that have $e$ as their common edge.

\subsection{Coarse-to-fine reconstruction}\label{subsec:ctf}
Once the virtual shape obtained from space carving has been optimized using Eq.~\eqref{eq:loss_function}, a new surface mesh is generated, which will serve as the initial shape for the next optimization stage. Before each stage, we remesh the surface~\cite{PTC10} in a coarse-to-fine fashion in order to recover progressively finer details. For remeshing, the target edge length $t_l$ is sampled from the inverse distance space, as defined below:
\begin{equation}
	\centering
	t_l = \frac{L \cdot t_{\textit{min}}}{l}, (l = 1, 2, ..., L),
	\label{eq:inv_space}
\end{equation}
where $t_{\textit{min}} = 0.005 \cdot \texttt{diaglen}$, and $\texttt{diaglen}$ is the length of the object's bounding box diagonal. Here $L$ is total number of optimization stages, which we set to 10 in our experiments.
The maximum distance between the surfaces before and after remeshing is set to $0.005\cdot\texttt{diaglen}$. %
During each optimization stage, for a single ray-pixel correspondence, we trace the camera ray and first locate the intersected triangles, whose associated vertices will be refined in this round of optimization. As the virtual shape becomes more accurate, it can more precisely filter out ray paths that involve more than two refractions, or total internal reflection, and provide a better initialization for subsequent optimization. The coarse-to-fine reconstruction of two models is visualized in Fig.~\ref{fig:bunny_kitten}.

\section{Results and Discussion}\label{sec:results}

\begin{figure}[t!]
	\centering
	\begin{overpic}
		[width=1.0\linewidth]{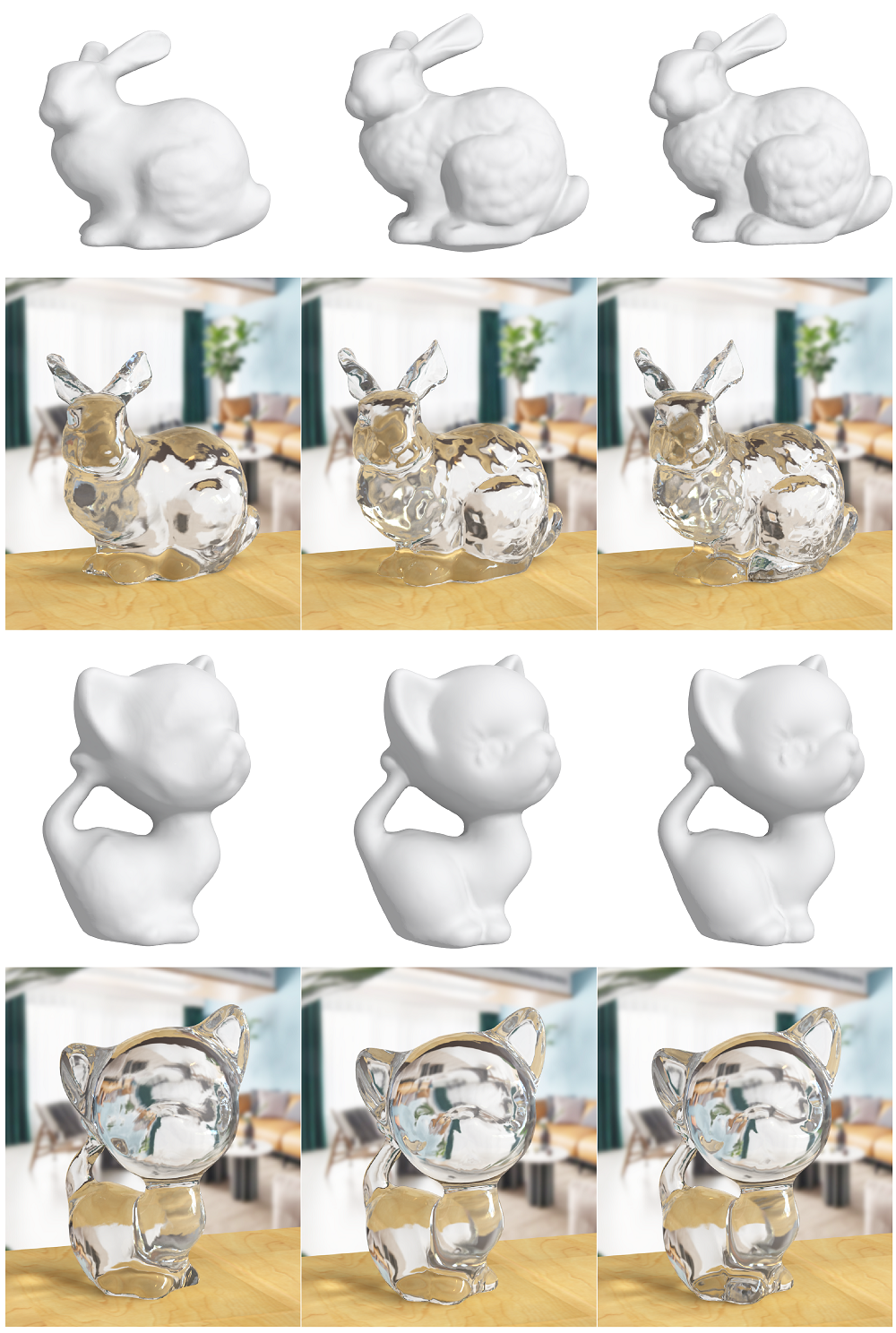}
		\put(5,-2){\small \cite{wu2018frt}}
		\put(31,-2){\small Ours}
		\put(50,-2){\small Ground Truth}
	\end{overpic}
	\caption{A qualitative comparison using rendering. The three images in each row are rendered from the same camera view, with the result of Wu \etal~\shortcite{wu2018frt} on the left, our reconstructed model in the middle, and the original model on the right. Our method captures more of the fine level geometric detail then that of Wu \etal, and the ray-traced images are nearly identical to those rendered from the ground truth models.}
	\label{fig:syn-render}
\end{figure}

\begin{figure}[t!]
	\centering
	\includegraphics[width=1.0\linewidth]{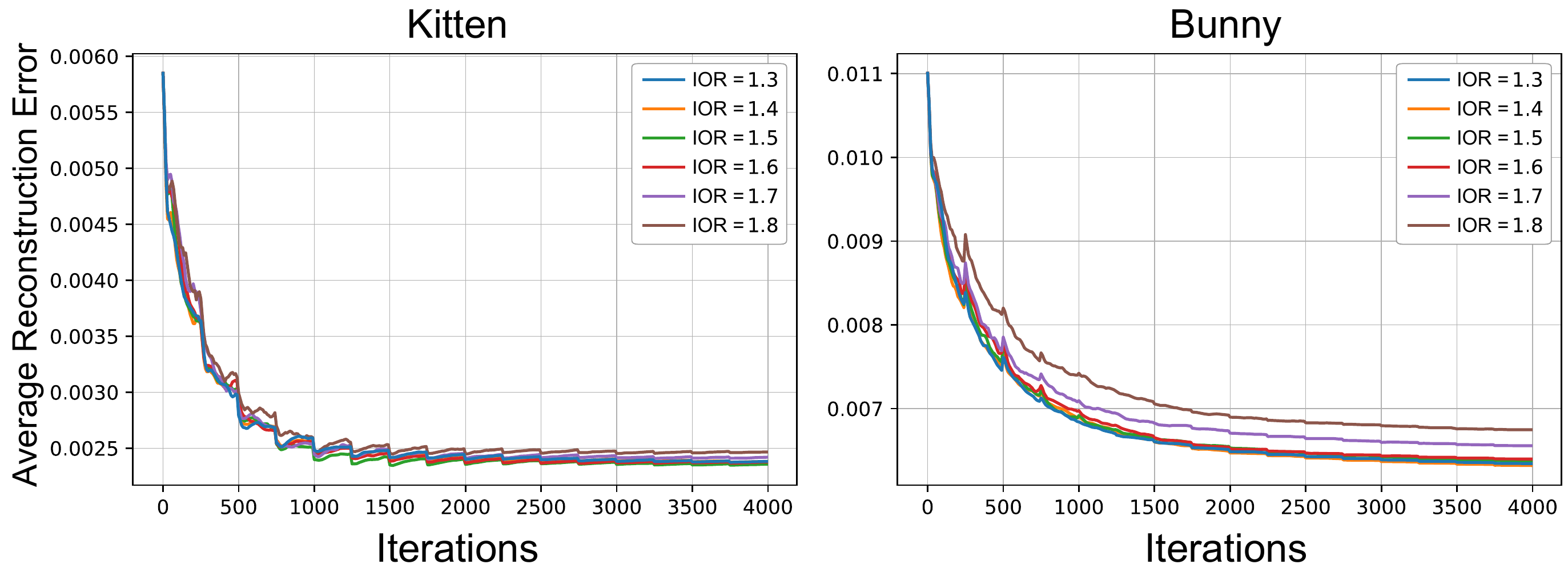}
	\caption{Reconstruction accuracy with different refractive indices. Errors are slightly higher for higher refractive indices, but the effect of the optimization is similar.}
	\label{fig:ior}
\end{figure}

\begin{figure}[t!]
	\centering
	\includegraphics[width=1.0\linewidth]{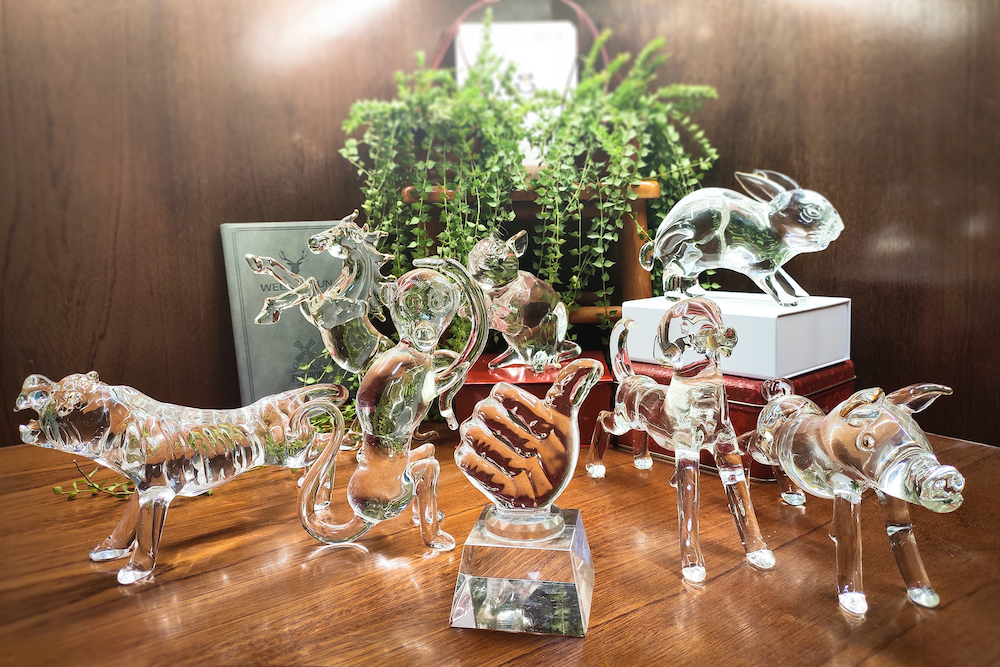}
	\caption{Real transparent objects used in our experiments. All of these objects exhibit geometric detail at a variety of scales.}
	\label{fig:objects}
\end{figure}

\begin{figure}[t!]
	\centering
	\includegraphics[width=1.0\linewidth]{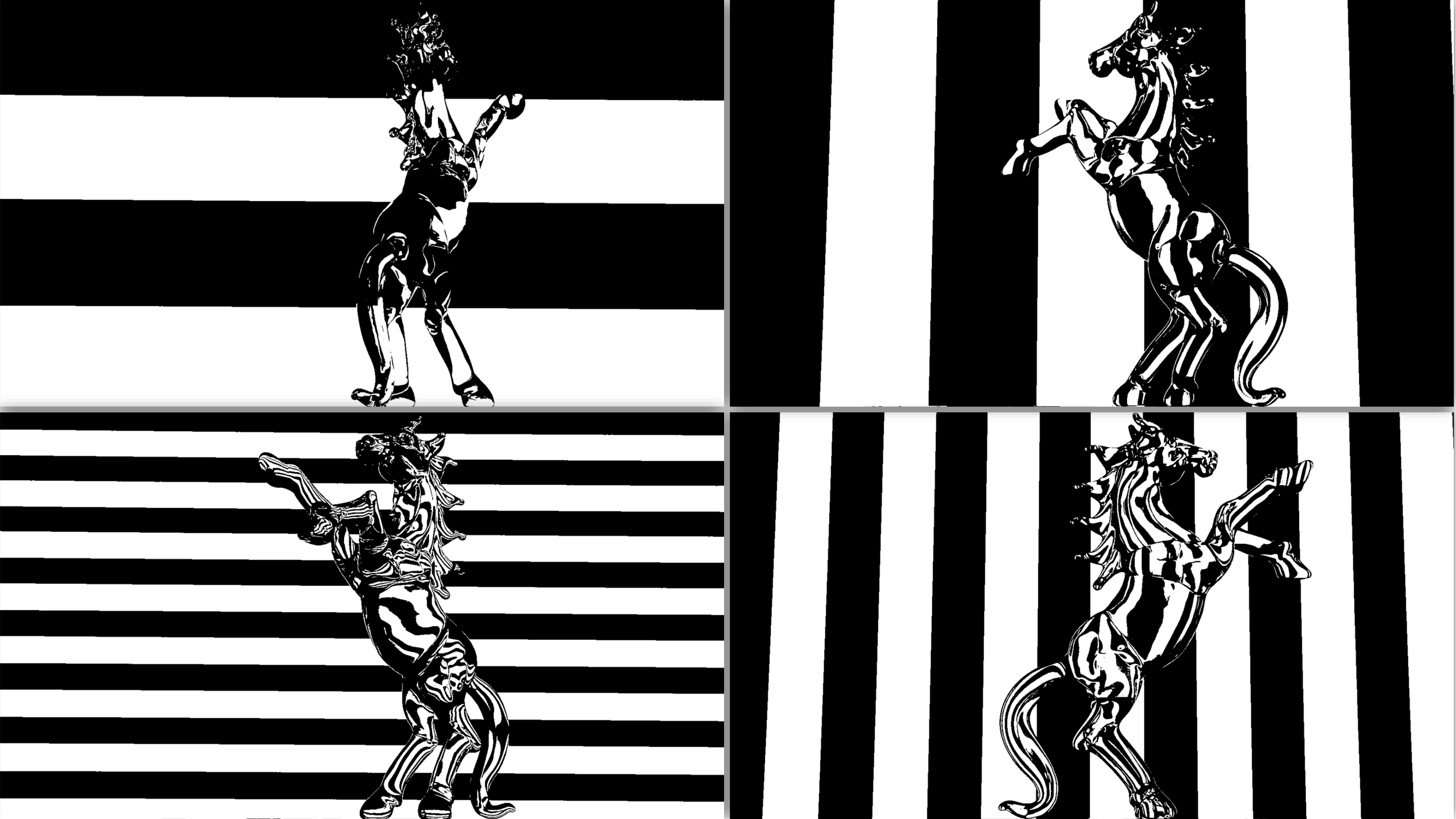}
	\caption{Four of the captured images of a real \textit{Horse} object, each image is taken from a different view and while the background monitor is displaying one of the horizontal or vertical stripe patterns. Acquiring each view with a full Gray-coded background pattern enables extracting the environment matte and the object silhouette.}
	\label{fig:real-data}
\end{figure}

\begin{figure*}[t!]
	\centering
	\includegraphics[width=0.94\linewidth]{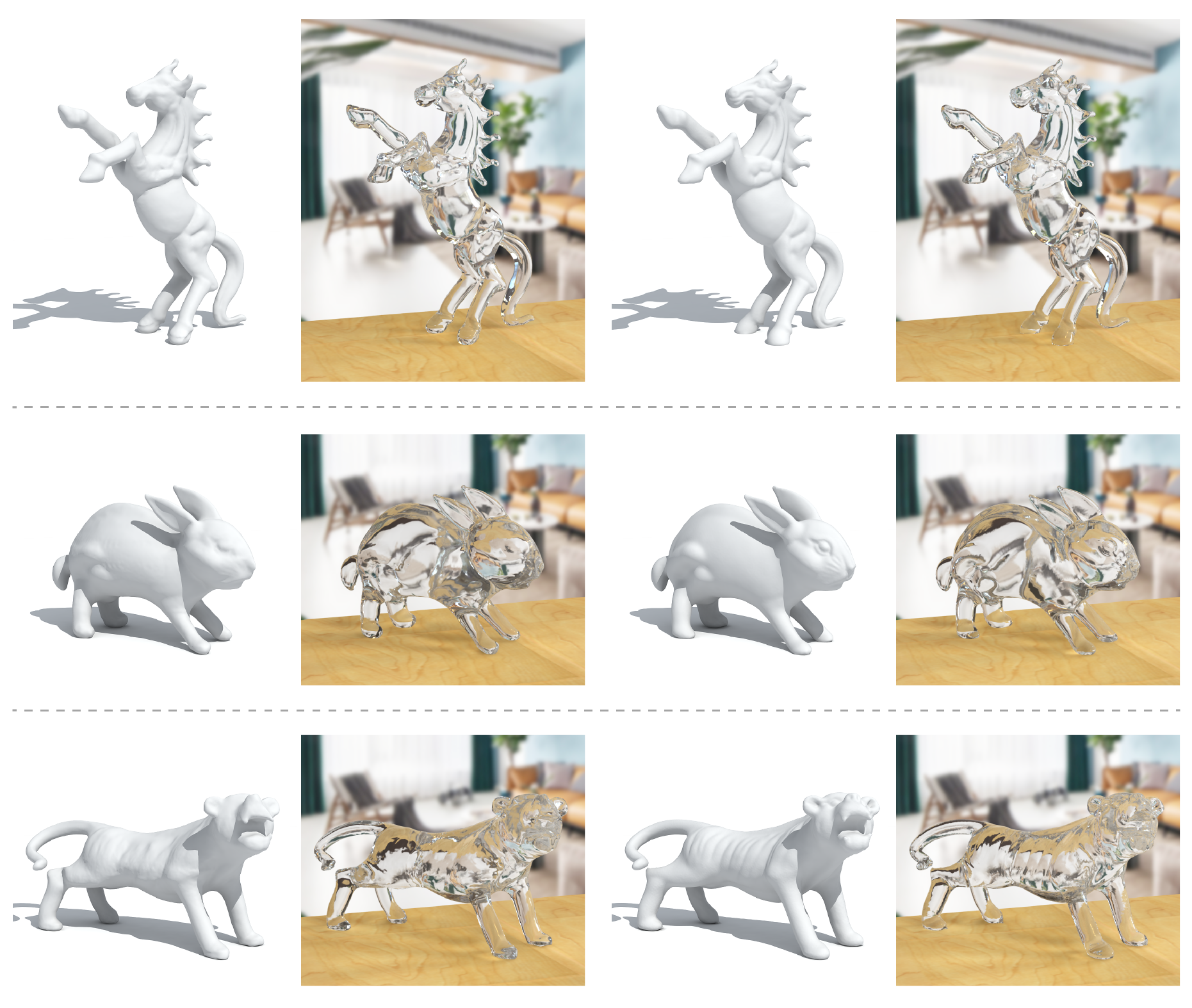}
	\caption{Reconstruction of real transparent objects: \textit{Horse}, \textit{Rabbit}, and \textit{Tiger}. For each object, our reconstructed model is rendered with diffuse shading and ray-traced in a virtual environment (two left columns). For qualitative comparison, the corresponding ground truth models are rendered in the two right columns. Note that our reconstructed models succeed in capturing some the skin folds and wrinkles on the body of \textit{Horse} and \textit{Tiger}, as well as the whiskers and eyes of \textit{Rabbit}.}
	\label{fig:results}
\end{figure*}

\begin{figure*}[t!]
	\centering
	\begin{overpic}
		[width=1.0\linewidth]{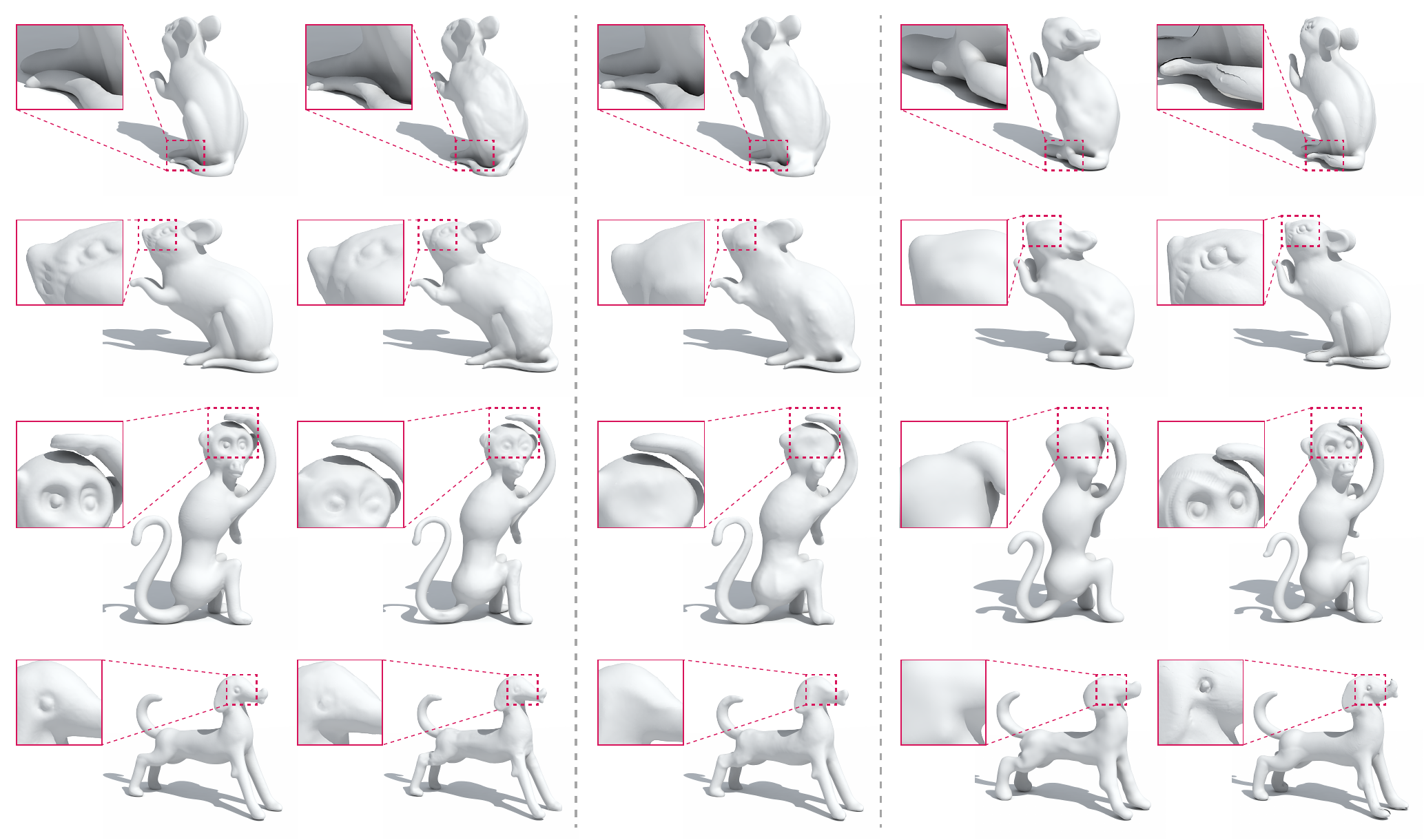}
		\put(5,-1){\small Our ground truth}
		\put(30,-1){\small Ours}
		\put(46,-1){\small \cite{wu2018frt}}
		\put(67,-1){\small \cite{li2020through}}
		\put(84,-1){\small Li~\etal's GT}
	\end{overpic}
	\caption{Comparison with Wu \etal~\shortcite{wu2018frt} and Li \etal~\shortcite{li2020through} using real transparent objects: \textit{Mouse}, \textit{Monkey} and \textit{Dog}. For each object, compared with its corresponding ground truth, our reconstructed results better capture geometric details at various scales, while the results of Wu \etal\ and Li \etal\ are over-smoothed and many of these details are lost. Note that, due to the lack of silhouette constraints, the arm of the \textit{Monkey} and the tail of the \textit{Mouse} in the reconstructions of Li \etal\ are falsely connected with the body, while our reconstruction of these thin structures is more precise.}
	\label{fig:comp}
\end{figure*}

\subsection{Acquisition}\label{subsec:setup}

We begin by briefly describing our acquisition setup. As shown in Fig.~\ref{fig:setup}, the transparent object to be captured is positioned on turntable, between an LCD monitor that displays the coded background pattern and a static camera, placed in front of the object. The intrinsic and extrinsic parameters of the camera, and the relative positions of monitor and turntable with respect to the camera are calibrated~\cite{Zhang_calib}, before the acquisition commences.

To capture an object, the turntable is rotated to a set of 72 evenly sampled viewing angles. \bj{At each viewing angle, a Gray-coded background pattern is displayed on the monitor for simultaneously extracting silhouette and estimating ray-pixel correspondences using environment matting. The Gray-coded background is produced by displaying a sequence of 11 images with vertical stripes and 11 images with horizontal stripes (see Fig.~\ref{fig:real-data}).}
Note that, in order to avoid the influence of ambient light, the entire acquisition process is conducted in a dark room, \bj{and the background monitor is used as the only light source.}

\subsection{Implementation details}\label{subsec:impl}

We use PyTorch~\cite{paszke2017automatic} to implement our approach. During each of the 10 optimization stages, the loss function is evaluated and its gradients are back-propagated 500 times. Each time, we randomly select one camera view for computing refraction loss and nine other views (spaced apart by $40^\circ$ from each other with a random starting view) for silhouette loss. 
To compute the refraction loss, we first use the OptiX ray tracing engine~\cite{parker2010optix} to find the triangles intersected by each ray path from the camera to the background.
\bj{The OptiX engine is used here for efficiency, which becomes important as the number of mesh triangles grows. However, OptiX is not a differentiable ray tracer, and thus, once the intersected triangles are known, the final intersection points ($p_1^\prime$ and $p_2^\prime$) are computed in PyTorch in a fully differentiable manner.
	Having computed these intersection points, we trace the ray as discussed in Section~\ref{subsec:loss_refract}, to obtain a differentiable ray path with regard to the mesh vertices.
} 
We then perform gradient descent using Nesterov momentum optimizer with default arguments, where the learning rate decays from $0.005 \cdot \texttt{diaglen}$ to $0.002 \cdot \texttt{diaglen}$. 
It takes about 30 minutes to perform the data acquisition for a single object, and about 1 minute to reconstruct the visual hull from the silhouettes using space carving. Starting from the visual hull, the progressive optimization takes about 10 minutes.

\subsection{Experiments with synthetic objects}
\label{subsec:synthetic}

We first evaluate our method on two synthetic mesh models: \textit{Bunny} and \textit{Kitten}, rendered as solid transparent objects with a refractive index of 1.5. To emulate the acquisition process, we render the models using a path tracer implemented using OptiX~\cite{parker2010optix}.
For photorealistic appearance, we also simulate total internal reflection, but limit the ray tracing depth to 30.
The virtual data capturing setup follows the one described in Section~\ref{subsec:setup}, with the camera set to a pinhole model, and the resolution of virtual background monitor set to $1920 \times 1080$.

As shown in Fig.~\ref{fig:bunny_kitten}, starting from the visual hull, the mesh is progressively refined and fine geometric details are gradually recovered. The distance between our intermediate reconstructed results and the ground truth models are measured using the minimal per-vertex distance and visualized using a colormap. The initial approximations exhibit large areas where the error is high (around 5 percent of the longest bounding box dimension), however these errors are significantly reduced as the optimization progresses. The average error also decreases, roughly to 50 percent of its starting value. 

Fig.~\ref{fig:syn-render} shows a qualitative evaluation of our results by rendering our final reconstructions (middle column) next to the ground truth (right column). For comparison, we also render the results of Wu \etal~\shortcite{wu2018frt} for these two synthetic models (left column). It may be seen that our results capture more fine level details, such as the bumps on the bunny's body or the eyes of the kitten, thereby yielding more accurate reconstructions. When simulating refractive light transport through each of these models, it may be seen that our reconstructions are accurate enough to closely approximate the images obtained with the original models. In contrast, the refractions through the models reconstructed by Wu \etal~\shortcite{wu2018frt} are visibly too smooth. 

Finally, in Fig.~\ref{fig:ior} we examine the accuracy of our reconstruction of the same two synthetic examples for different refractive indices. A higher refractive index, induces stronger changes in the direction of the refracted light rays, which may cause more of the final exiting rays to miss the background monitor. This leads to fewer valid ray-pixel correspondences and causes a slight decrease in the reconstruction accuracy. The local peaks in the error plots in Fig.~\ref{fig:ior} occur at 500 iteration cycles are caused by remeshing, which temporarily increases the reconstruction error.

\begin{figure}[t!]
	\centering
	\begin{overpic}
		[width=1.0\linewidth]{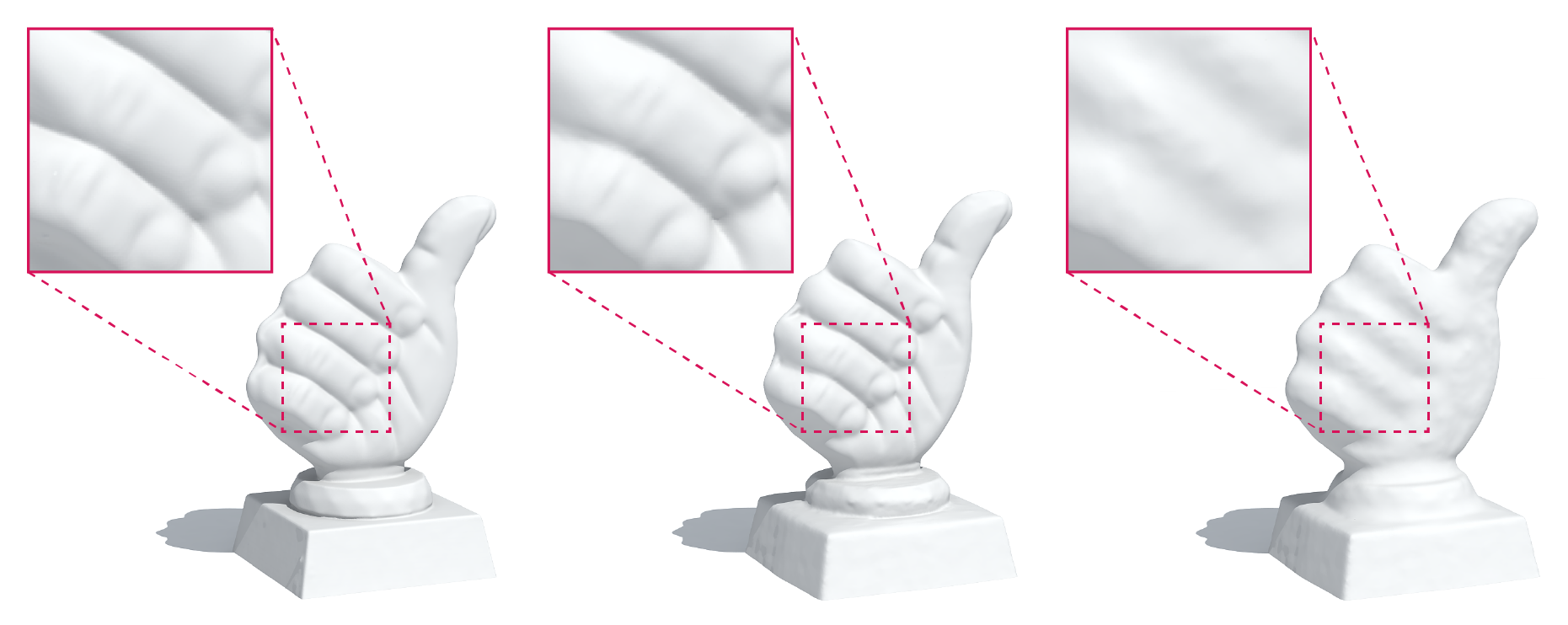}
		\put(10,-2){\small Ground truth}
		\put(50,-2){\small Ours}
		\put(76,-2){\small \cite{wu2018frt}}
	\end{overpic}
	\caption{Comparison with Wu \etal~\shortcite{wu2018frt} using a real \textit{Hand} object. Our result succeeds in capturing the finger nails, and the creases between fingers, while these details are smoothed over by the method of Wu \etal}
	\label{fig:comp-hand}
\end{figure}

\begin{figure*}[t!]
	\centering
	\begin{overpic}
		[width=1.0\linewidth]{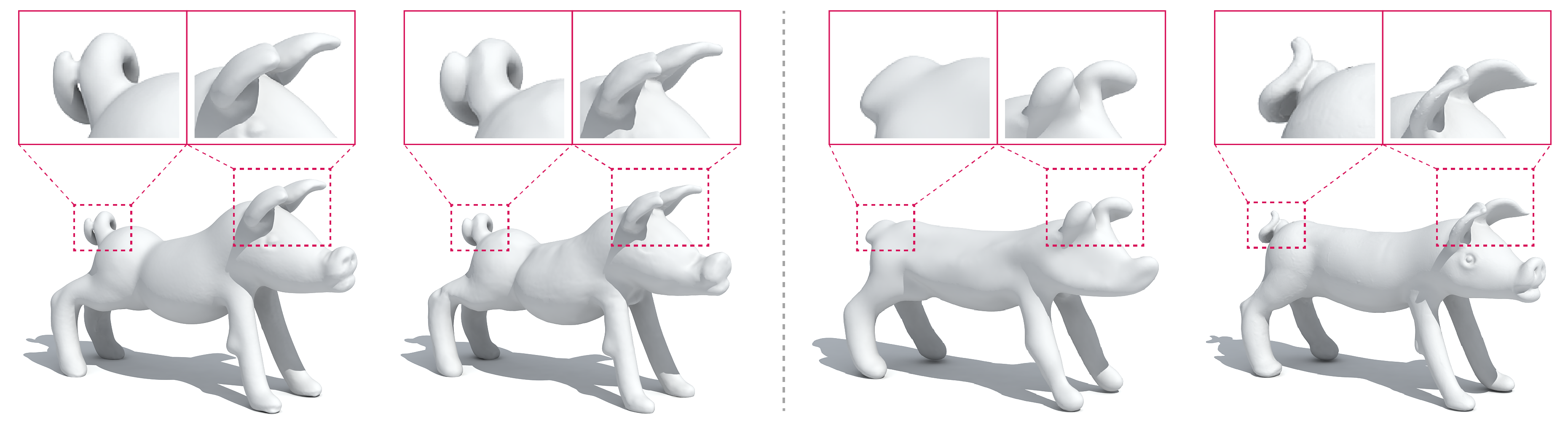}
		\put(5,-1){\small Our ground truth}
		\put(35,-1){\small Ours }
		\put(58,-1){\small \cite{li2020through}}
		\put(82,-1){\small Li~\etal's GT}
	\end{overpic}
	\caption{Comparison with Li \etal~\shortcite{li2020through} using a real \textit{Pig} object. Fine details on the body, such as the eye, are smoothed over, and the shape of the ears and tail are distorted in their result,
		in contrast to our method.}
	\label{fig:comp-pig}
\end{figure*}

\begin{figure}
	\centering
	\begin{overpic}
		[width=1.0\linewidth]{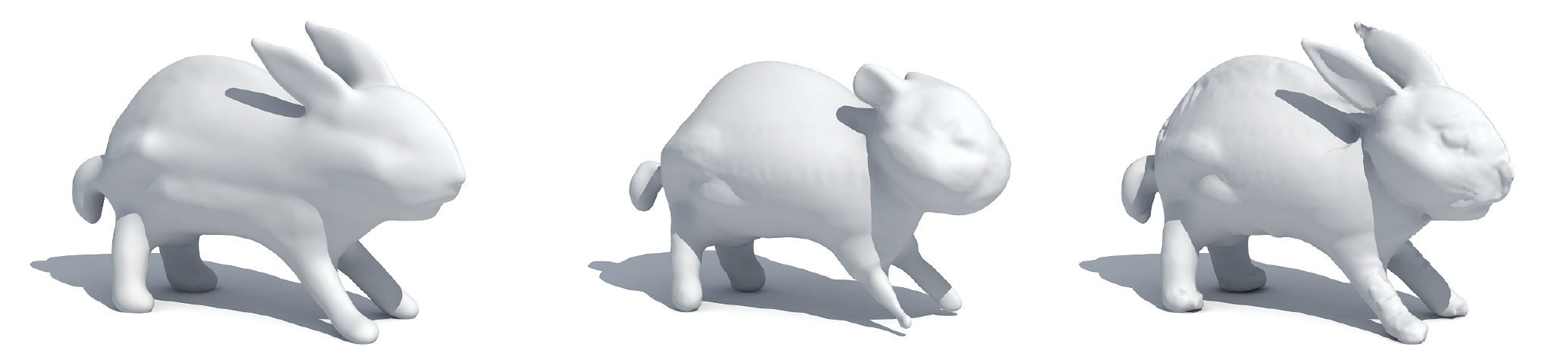}
		\put(2,-3){\small w/o refraction loss}
		\put(37,-3){\small w/o silhouette loss}
		\put(68,-3){\small w/o smoothness loss}
	\end{overpic}
	\caption{Reconstruction of \textit{Rabbit} without different loss terms. The full reconstruction result is shown in Fig.~\ref{fig:results}, where the average reconstruction error is 0.6261mm. The errors for the above three reconstructions (from left to right) are 0.8420mm, 2.2337mm and 0.7300mm, respectively.}
	\label{fig:ablation_loss}
\end{figure}

\begin{figure}
	\centering
	\begin{overpic}
		[width=1.0\linewidth]{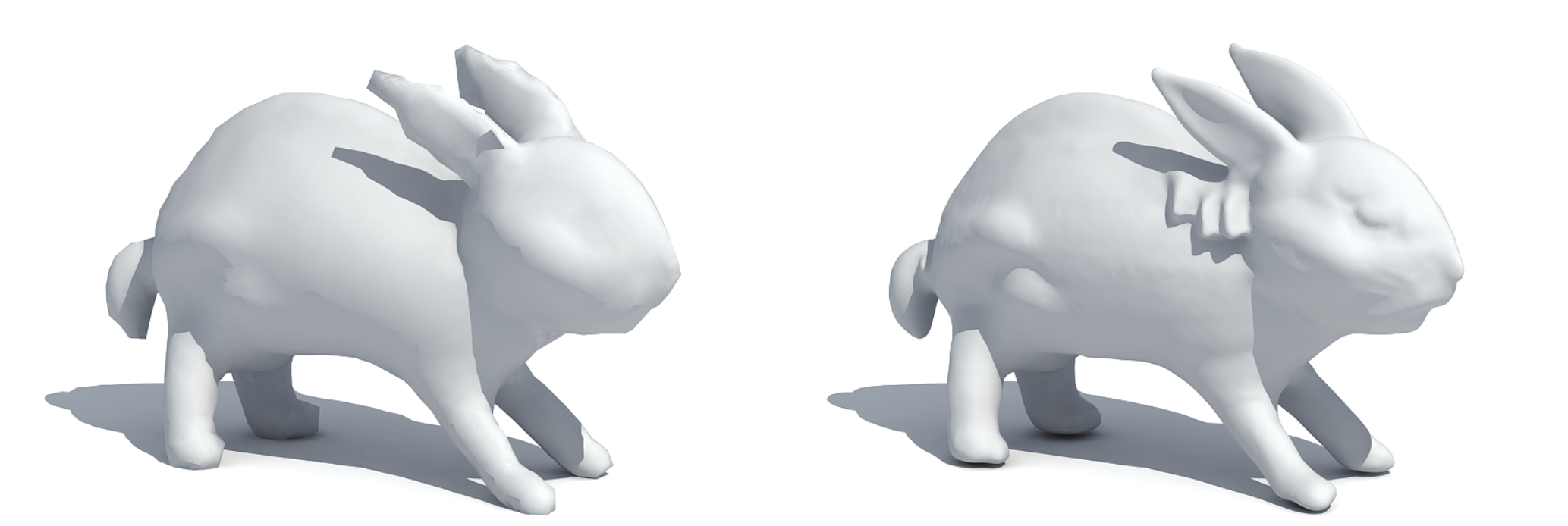}
		\put(5,-3){\small Coarse-level optimization}
		\put(57,-3){\small Fine-level optimization}
	\end{overpic}
	\caption{Validation of our coarse-to-fine optimization strategy. The left reconstruction is achieved by optimizing directly the initial coarse mesh without further remeshing, while the right reconstruction directly optimizes the final fine-level mesh. The average reconstruction errors are 0.7376mm and 0.9187mm, respectively, compared to an average error of 0.6261mm for the coarse-to-fine reconstruction.}
	\label{fig:ablation_ctf}
\end{figure}

\subsection{Experiments with real objects}\label{subsec:real_exp}

We obtained eight real transparent objects made from borosilicate 3.3 glass (the refractive index is 1.4723) for our experiments; see Fig.~\ref{fig:objects}. For data acquisition, a DELL LCD monitor with resolution $1920 \times 1200$ is used to display the Gray-coded background patterns, serving as the only illumination source. The turntable is controlled by a high precision stepper motor to rotate $5^\circ$ each time. A single PointGrey Flea3 color camera is used to capture the images. We used an aperture of f/6.0, 
while the shutter time is set to about 50ms for adequate exposure; see Fig.~\ref{fig:real-data} for an example of acquired data.

In order to quantitatively evaluate the accuracy of reconstruction, we paint each object with DPT-5 developer and then scan it with a high-end industrial level scanner\footnote{Artec Eva, https://www.artec3d.com/portable-3d-scanners/artec-eva} to obtain a ground truth 3D mesh. We compare between the reconstructed results and the ground truth after aligning them using ICP, as shown in Fig.~\ref{fig:lod} and~\ref{fig:hand}. Quantitatively, our reconstructed models improve the approximation provided by the visual hull roughly by a factor of x2, and the average distance of our final reconstructed models from the ground truth is on the order of 0.1-0.3 percent of the model's bounding box diagonal. The average reconstruction errors for all eight real objects are reported in Table~\ref{tab:are}, where we also compare our accuracy with the state-of-the-art.

Qualitatively, our final reconstructions is able to noticeably improve upon the initial visual hull model, while maintaining the accuracy of the silhouettes, even for thin structures such as the tail in Fig.~\ref{fig:lod}.
Additional examples can be found in Fig.~\ref{fig:results}, where the reconstructed models of the \textit{Horse}, \textit{Rabbit} and \textit{Tiger} objects, demonstrate the ability of our method to cope with various complicated geometries, capturing fine scale geometric details, such as skin folds, wrinkles, whiskers, etc.

\begin{table}[t!]
	\centering
	\caption{Reconstruction error, measured by average per-vertex distance, of our results,~\cite{wu2018frt} and~\cite{li2020through}, each with respect to its corresponding ground truth. As a baseline, we also report the average distance between the initial visual hull and the ground truth. All distances are normalized by the bounding box diagonals of the corresponding models.}
	\resizebox{\linewidth}{!} {
		\begin{tabular}{c|cccc}
			\hline
			\textbf{}       & initial  & \cite{wu2018frt}    & \cite{li2020through}   & ours      \\ \hline
			\textit{Mouse}  & 0.007164 & 0.005544 & 0.005840 & \textbf{0.003075} \\
			\textit{Dog}    & 0.004481 & 0.002678 & 0.002778 & \textbf{0.002065} \\
			\textit{Monkey} & 0.005048 & 0.003011 & 0.004632 & \textbf{0.002244} \\
			\textit{Hand}   & 0.005001 & 0.005170 & -        & \textbf{0.002340} \\
			\textit{Pig}    & 0.004980 & -        & 0.004741 & \textbf{0.002696} \\
			\textit{Rabbit} & 0.005639 & -        & -        & \textbf{0.002848} \\
			\textit{Horse}  & 0.002032 & -        & -        & \textbf{0.001160} \\
			\textit{Tiger}  & 0.005364 & -        & -        & \textbf{0.003020} \\ \hline
		\end{tabular}
	}
	\label{tab:are}
\end{table}

\begin{figure*}
	\centering
	\begin{overpic}
		[width=1.0\linewidth]{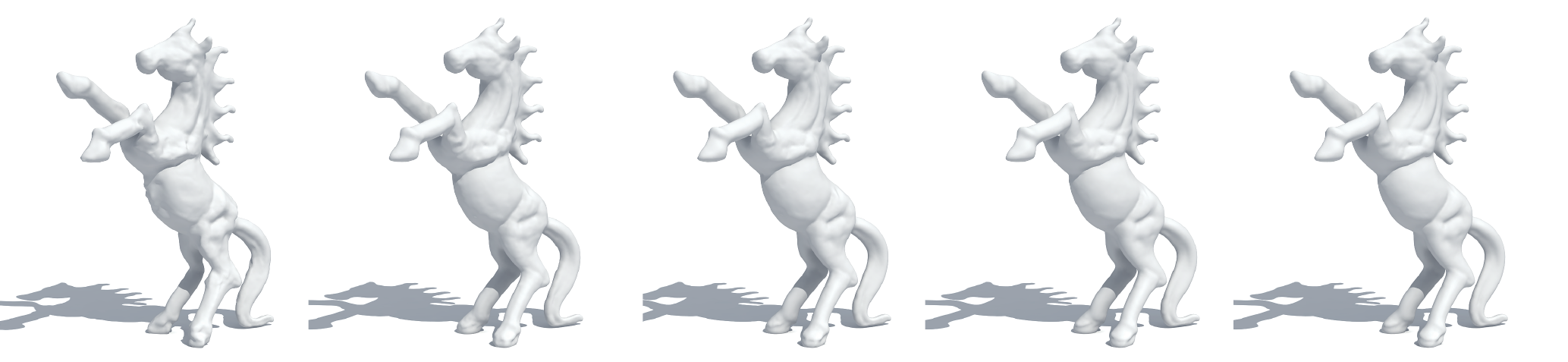}
		\put(8,-1){\small 4 views}
		\put(28,-1){\small 9 views}
		\put(48,-1){\small 18 views}
		\put(68,-1){\small 36 views}
		\put(88,-1){\small 72 views}	
	\end{overpic}
	\caption{Optimization using different numbers of views. In each case, the views directions are evenly sampled from $360^\circ$. As the number of views increases, the reconstruction results gradually capture more geometric details. The average reconstruction errors are plotted in Fig.~\ref{fig:simulated_real} (left, solid green line).}
	\label{fig:ablation_views}
\end{figure*}

\begin{figure*}
	\centering
	\includegraphics[width=1.0\linewidth]{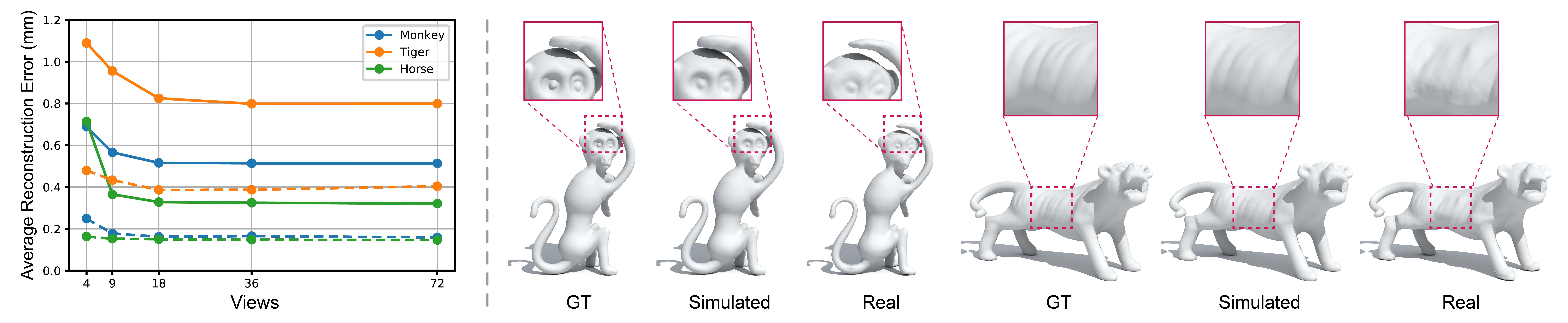}
	\caption{\bj{Average reconstruction error (in millimeters) vs.~the number of views used by the optimization for \textit{Monkey}, \textit{Tiger} and \textit{Horse}. The solid lines represent experiments with real objects, as described in Section~\ref{subsec:real_exp}. For comparison, the dotted lines represent experiments with synthetic objects (obtained by scanning the real ones), as described in Section~\ref{subsec:synthetic}. 
	In each case, compared with the ground truth, the simulated experiment with more accurate ray-pixel correspondences clearly demonstrate higher quality over our real results. All reconstructions were obtained by optimizing over the full set of 72 views. }}
	\label{fig:simulated_real}
\end{figure*}

\subsection{Comparisons with state-of-the-art}

The relevant state-of-the-art methods are those of Wu \etal~\shortcite{wu2018frt} and Li \etal~\shortcite{li2020through}, since both of them reconstruct the full shape of a transparent object in a non-intrusive way, as described in Section~\ref{sec:related}.
To perform the comparison with Wu \etal~\shortcite{wu2018frt}, we obtained from them their captured data, their reconstructed models, and the corresponding ground truth scans. Similarly, we obtained the reconstructed and the ground truth models from Li~\etal~\shortcite{li2020through}. It should be noted that although Li~\etal\ experimented with transparent objects obtained from the same source, \bj{due to different manufacturing batches} there are slight differences between the shapes, and therefore their results are compared to a different set of ground truth models.

As demonstrated in Fig.~\ref{fig:comp}, the reconstructions produced by either of these two state-of-the-art methods tend to smooth out the fine geometric details. In contrast, our method is more successful in capturing geometric detail at various scales, such as the larger scale displacements of the neck and tummy for the \textit{Mouse} and \textit{Monkey} objects, as well as the smaller scale details of the eyes of \textit{Monkey} and \textit{Dog}. Furthermore, the silhouettes for thin structures, such as the tail of \textit{Mouse} or the arm of \textit{Monkey}, are better preserved. Additional comparisons may be found in Fig.~\ref{fig:comp-hand} and Fig.~\ref{fig:comp-pig}, where in each case, the reconstructions generated by our method are visibly closer to the ground truth.

\begin{figure*}
	\centering
	\includegraphics[width=1.0\linewidth]{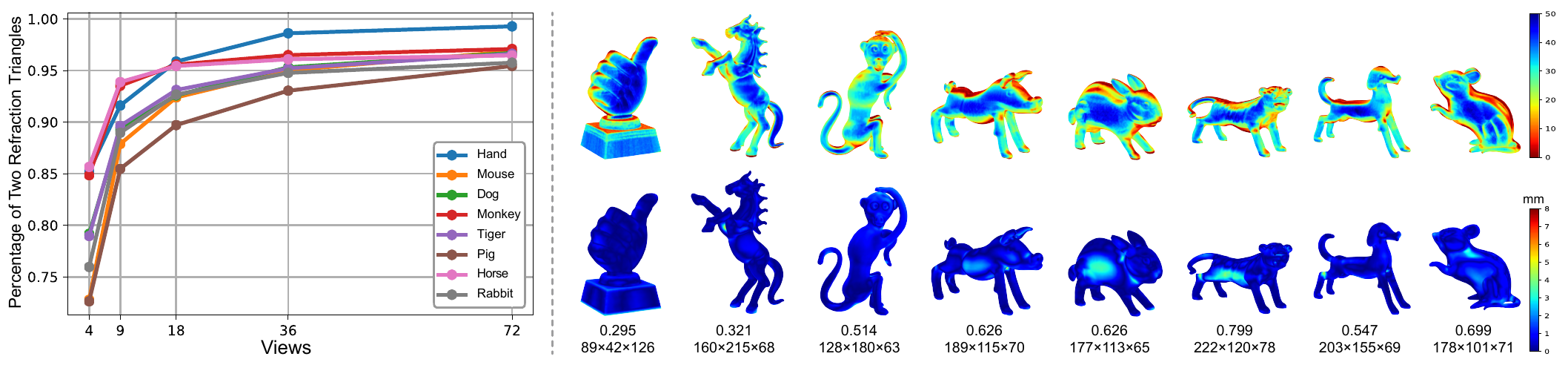}
	\caption{\bj{Left: The percentage of the final mesh triangles that are pierced by at least one two-refraction ray path vs. the number of views. Right top: visualization of the number of two-refraction paths piercing each triangle under 72 views. The maximum number is truncated to 50 for better visualization. Right bottom: the corresponding reconstruction error map. Below each model we report the average reconstruction error and the real size of bounding box, in millimeters.}}
	\label{fig:tris_vis}
\end{figure*}

\begin{figure}
	\centering
	\includegraphics[width=1.0\linewidth]{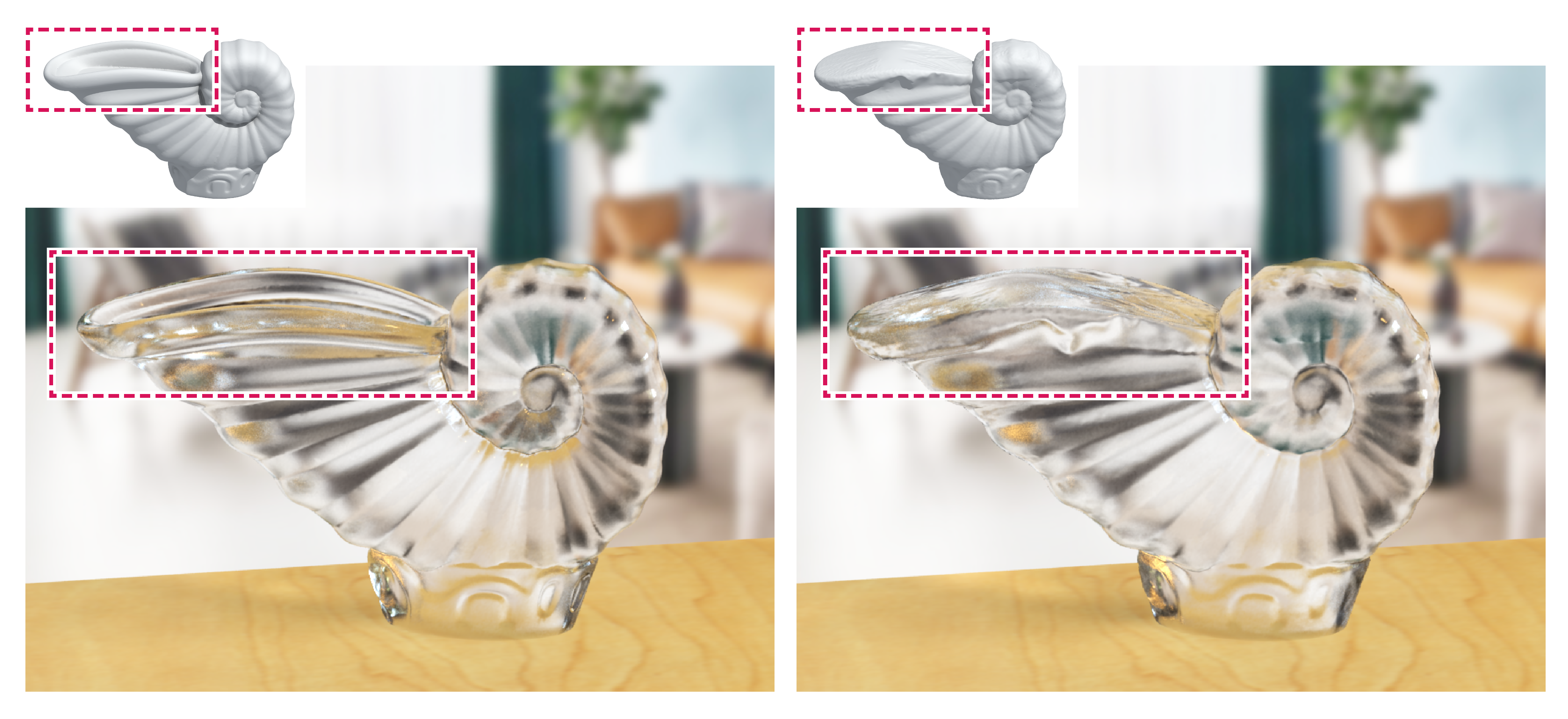}
	\caption{\bj{Failure case. Left: a \textit{Shell} model with a cavity (highlighted by the dotted rectangle) and its rendering. Right: the result of reconstruction and rendering for a qualitative comparison. Because of multiple refractions in the highlighted region, the cavity fails to be reconstructed correctly.}}
	\label{fig:failure}
\end{figure}

\subsection{Ablation study}\label{subsec:ablation}

We first validate the effectiveness of each component of our loss function in Eq.~\eqref{eq:loss_function}, using the \textit{Rabbit} object. The reconstruction using the full loss function is shown in the rightmost column of Fig.~\ref{fig:results}, while reconstructions corresponding to omissions of each of the three loss terms are shown in Fig.~\ref{fig:ablation_loss}.
It may be seen that when the refraction loss $\Lref$ is omitted, the reconstruction fails to capture the finer scale geometric details, which are captured by the refraction loss since they affect the refractive light transport through the object. We conclude that the refraction loss plays a key role in the recovery of fine details.
On the other hand, omitting the silhouette loss $\Lsil$ fails to correctly reproduce the overall shape, which is particularly noticeable in the ears, legs, and tail of the rabbit. Finally, omitting the smoothness loss $\Lsm$ results in macroscopic roughness artifacts (on legs and back).

Next, Fig.~\ref{fig:ablation_ctf} examines the effect of our coarse-to-fine optimization strategy. Specifically, we show the results when optimizing the initial coarse-level mesh (left) and the final fine-level mesh (right), without progressing from one to another by remeshing every 500 iterations. It is not surprising that optimizing the coarse-level mesh alone, cannot recover the finer surface details, due to its sparse sampling of the shape. However, optimizing the fine-level mesh directly results in some displacement artifacts, since the optimization is prone to getting locally stuck in local minima. 

We examine the effect of the number of views on reconstruction accuracy in Figs.~\ref{fig:ablation_views} and \ref{fig:simulated_real}. As more views are incorporated into the optimization, the number of valid ray-pixel correspondences that may be used by the optimization is increased as well. \bj{This increase, in turn, improves the reconstruction accuracy, as reported for 3 real objects in Fig.~\ref{fig:simulated_real} (left, plotted in solid lines). Nonetheless, note that the accuracy gains diminish as more and more views are added, and acceptable results may be obtained with 18 or 36 views.}

\bj{\subsection{Discussion}\label{subsec:discussion}

To further examine the effect of the two-refraction assumption, for each of our real examples, we measure the percentage of the final mesh triangles that are pierced by at least one two-refraction ray path, when considering all of the captured views. Fig.~\ref{fig:tris_vis} (left) shows that as the number of views increases, the average percentage of such triangles goes up (from 70.5\% to 92.4\%).
Thus, when using 36 uniformly spaced horizontal rotation angles, more than $90\%$ of the triangles (on average across all eight examples) are being optimized using two-refraction paths. The corresponding increase in the reconstruction accuracy is reported in Fig.~\ref{fig:simulated_real} (left, solid lines). We also visualize the number of two-refraction paths piercing each triangle with 72 views and the corresponding reconstruction error map of each object on the right side of Fig.~\ref{fig:tris_vis}. In general, the number of two-refraction paths tends to be smaller for those triangles whose normals are nearly perpendicular to the camera view rays (for all views). On the other hand, those parts of the model are typically well captured by the silhouettes.
Thus, it could be argued that the refraction loss complements the visual hull and the silhouette loss, and the deficiency of two-refraction paths does not result in larger errors.
In fact, Fig.~\ref{fig:tris_vis} shows that the largest errors in the \textit{Rabbit} and \textit{Tiger} models occur on the sides, which receive plenty of two-refraction paths. The reason is that the initial visual hull in these areas happens to deviate significantly from the true shape, and the silhouette loss is not able to fix it completely. 
}

We have demonstrated the ability of our method to reconstruct accurate and detailed models of transparent objects; however, several sources of errors remain. \bj{In order to better understand them, in Fig.~\ref{fig:simulated_real} we also plot the reconstruction accuracy for the synthetic counterparts of the eight real objects, where the reconstruction is carried out as described in Section~\ref{subsec:synthetic}. The results, plotted using dotted lines in Fig.~\ref{fig:simulated_real}, show that the accuracy is significantly improved. Reconstructing the synthetic models effectively eliminates all the real-world factors that could introduce errors, since the rendered images are noise-free and the ray-pixel correspondences are calculated precisely. As shown in Fig.~\ref{fig:simulated_real}, the eye of \textit{Monkey} and the side of \textit{Tiger} are reconstructed better than for the real objects. This suggests that} one source of error is the uncertainty in the data acquisition, and other possible errors in the environment matte extraction process. Another limitation on the reconstruction accuracy is imposed by the limited size of the background monitor. Due to this limited size, some of the ray paths never reach the coded background, and fewer valid ray paths are thus available to optimize the shape with. This results in a reduction of the reconstruction accuracy, as was already shown in Fig.~\ref{fig:ior}.

Furthermore, \bj{as shown in Fig.~\ref{fig:simulated_real}, the accuracy curves tend to flatten out after 18 views in almost all cases. We believe the reason is that at that point we have sufficiently many twice refracted ray paths and silhouetted, thus the errors begin to be dominated by other factors, such as considering only pure refractions and ignoring reflections,} which is not enough for faithful simulation of light transport through transparent objects. Thus, we assume a one-to-one correspondence between camera rays and points on the background, while in reality multiple background locations may contribute to a single image pixel.

\bj{Fig.~\ref{fig:failure} demonstrates a failure case of our method. Here the \textit{Shell} reconstructed object has a hollow part near the opening (highlighted in the figure). Because all of the views are horizontal, this cavity cannot be captured by the object's silhouettes, and all of the ray paths passing through that part (from all of the views) require more than two refractions in order to reach the background. Thus, our approach fails to correctly reconstruct the cavity, as shown in Fig.~\ref{fig:failure}.}

\section{Conclusions}
\label{sec:conclusion}

Although several dedicated approaches for recovering the 3D shape of a transparent object have been recently proposed, 
these methods are still unable to approach the accuracy of 3D scanning for opaque diffuse objects.
In this work, we have proposed a new transparent object reconstruction method, which delivers a significant improvement in reconstruction accuracy and its ability to capture fine geometric surface details.
Our controlled capture setup is considerably simpler than that of Wu \etal~\shortcite{wu2018frt}, and the reconstruction process is more streamlined.
Both of these significant improvements may be attributed to the power of differentiable ray tracing, which we use to directly optimize the recovered 3D mesh so as to minimize a combination of several complementing loss terms.

The main limitation of our approach is that only rays that undergo two refractions through the object are taken into account by our optimization process.
Future work should attempt to alleviate this restriction, as well as extend our approach to more general optical properties, such as objects which might not be homogeneous in their color, refractive index, or transmission properties.

\begin{acks}
We sincerely thank Zhengqin Li, Yu-ying Yeh and Manmohan Chandraker for providing us their reconstructed results and their scanned ground truth of \textit{Mouse}, \textit{Monkey}, \textit{Dog}, and \textit{Pig} as shown in~\cite{li2020through}. We also thank the anonymous reviewers for their valuable comments. This work was supported in parts by NSFC (61761146002, 61861130365), GD Science \& Technology Program (2020A0505100064, 2018KZDXM058, 2018A030310441, 2015A030312015), GD Talent Plan (2019JC05X328), LHTD (20170003), Israel Science Foundation (2366/16, 2472/17), National Engineering Laboratory for Big Data System Computing Technology, and Guangdong Laboratory of Artificial Intelligence and Digital Economy (SZ).
\end{acks}

\bibliographystyle{ACM-Reference-Format}
\bibliography{DRT}

\end{document}